\documentclass[prd,onecolumn,nofootinbib,superscriptaddress,floatfix,floats,11pt]{revtex4-2} 

\usepackage{graphicx}
\usepackage{amsfonts}
\usepackage{amssymb}
\usepackage{amsbsy}
\usepackage{amsmath}
\usepackage{mathrsfs}
\usepackage{latexsym}
\usepackage{natbib}
\usepackage{bm}
\usepackage[utf8]{inputenc}
\usepackage{subfigure} 
\usepackage{color}
\usepackage[active]{srcltx}
\usepackage{wasysym}
\usepackage{mathbbol}
\usepackage{bigints}
\usepackage{hyperref}
\allowdisplaybreaks
\usepackage{comment}
\usepackage{empheq}
\usepackage{soul}
\usepackage[export]{adjustbox}
\usepackage{gensymb}
\usepackage[utf8]{inputenc}
\newcommand{\bea}[1]{\begin{eqnarray}{#1}}
\newcommand{\eea}{\end{eqnarray}}
\newcommand{\dd}{\mathrm{d}}
\newcommand{\EE}{\mathrm{E}}
\newcommand{\LL}{\mathrm{L}}
\newcommand{\tr}{\mathrm{tr}}

\newcommand{\im}{\mathrm{Im}}

\reversemarginpar

\begin{document}

\title{Effective Action and Gravitational Pair Production in (A)dS Spacetime}

 \author{Yu Zhou}
 \email{yuzhou\_@buaa.edu.cn}
 \affiliation{Center for Gravitational Physics, Department of Space Science, Beihang University, Beijing 100191, China}
 \author{Hai-Qing Zhang}
 \email{hqzhang@buaa.edu.cn}
 \affiliation{Center for Gravitational Physics, Department of Space Science, Beihang University, Beijing 100191, China}
 \affiliation{Peng Huanwu Collaborative Center for Research and Education, Beihang University, Beijing
100191, China}

\begin{abstract}
\noindent
We compute the effective action for a massive scalar field in (A)dS spacetime using the Euclidean heat kernel method. We highlight that in even-dimensional dS spacetimes, the effective action exhibits a non-trivial imaginary part, reminiscent of the Schwinger effect in quantum electrodynamics.  We find consistency between the results obtained from the Euclidean heat kernel method with those from the Green's function approach in Lorentzian signature. Additionally, we compare our results with the perturbative calculations and find that the perturbation theory almost fails to capture the correct non-perturbative imaginary part of the effective action. This discrepancy presents a challenge to computing the gravitational pair production using the perturbation theory.
\end{abstract}

\maketitle
\tableofcontents
\section{Introduction}
In the realm of theoretical physics, particle production in the absence of perturbative frameworks reveals some intriguing phenomena in quantum field theory (QFT) and cosmology. Two notable examples of non-perturbative particle production are the Hawking effect \cite{Hawking:1975vcx} and the Schwinger effect \cite{PhysRev.82.664}. 

The Hawking effect describes the process in which black holes can emit particles and lose mass. This effect results in the evaporation of the black hole, which challenges our classical understanding of these enigmatic objects and raises profound questions about the fate of information in the black holes \cite{Page:2004xp}.
According to the QFT, vacuum is teeming with virtual particles constantly popping in and out. Hawking's heuristic picture indicates that near the event horizon of a black hole, because of the gravity, these virtual particles can become real, with one particle falling into the black hole and the other escaping out.
A similar effect of spontaneous production of particles from vacuum was discovered earlier in the framework of quantum electrodynamics (QED), known as Schwinger effect, which predicts the production of charged particle pairs in the presence of a strong electric field. According to the standard interpretations, the particles of a spontaneously created virtual pair are accelerated in the opposite directions by the external field. If the virtual particles gain enough energy over the distance of a Compton wavelength and obey the relativistic energy-momentum relation, they will become real particles \cite{PhysRevLett.130.221502}.
In Hawking's picture, the gravitational field plays the role of the electric field in the Schwinger effect. The similarity between the Hawking's picture and the Schwinger's picture naturally motivates us to investigate whether this kind of gravitational pair production can be explained by the Schwinger mechanism as well.

As described in the standard QFT textbooks, the signal of particle production can be extracted from the vacuum-to-vacuum scattering amplitude (vacuum persistence amplitude) \cite{Schwartz_2013},
\bea{}\label{defWeff1}
\left \langle \text{Out}  | \text{In}  \right \rangle =\exp\left(i W[g,A] \right) = \int \mathcal{D}[\phi]e^{i S[\phi,g,A]},
\eea
where $S[\phi,g,A]$ is the classical action of the field $\phi$ in external electromagnetic or gravitational fields while $W[g,A]$ is called the effective action. The amplitude determines the probability that the In-vacuum transits to Out-vacuum, with the Out-vacuum being still a vacuum and not containing In-particles. When the effective action $W$ is real, the transition probability from the In- to the Out- vacuum is equal to one. However, if the effective action has an imaginary part the probability of such a transition is not equal to one,
\bea{}
|\left \langle \text{Out}  | \text{In}  \right \rangle|^2=\exp\left(-2\text{Im} W \right).
\eea
This means that there is a probability of $P =1-|\left \langle \text{Out}  | \text{In}  \right \rangle|^2 \approx 2\text{Im} W $ for some pairs of particles to be produced. Therefore, the non-zero imaginary part of the effective action characterizes the production of particles.

In QED, the effective action is expressed by the so-called Euler–Heisenberg (EH) Lagrangian, see Ref.\cite{Dunne:2004nc} for review. In a uniform electric field $E$, the EH Lagrangian reads \cite{Schwartz_2013},
\bea{}
\mathcal{L}_{\text{EH}}=\frac{1}{2}E^2-\frac{1}{8\pi^2}\int_0^{+\infty}\frac{\dd s}{s^3}e^{i\epsilon s}e^{-m^2s}\left[ eEs\cot\left(eEs\right)-1+\frac{1}{3}\left( eEs\right)^2\right],
\eea
where $\epsilon>0 $ is an infinitesimal parameter.  In this form, we can see that, for real $E$, the integrand in the Lagrangian has poles in the range of the integration as $s$ equals $s_n=\frac{n\pi}{eE}, (n=1,2,\cdots)$. One needs to collect all these poles to obtain the non-zero imaginary part of the effective Lagrangian as, 
\bea{}\label{SchwingerEffect}
\im \mathcal{L}_{\text{EH}}=\frac{1}{8\pi}\sum_{n=1}^{\infty}\frac{1}{s_n^2}e^{-m^2s_n}.
\eea
This indicates that the strong electric fields can create electron–positron pairs, which is the Schwinger process. 

However, for generic systems, such as quantum fields in a classical black hole spacetime, it is difficult to obtain a closed form of the effective action. For this reason, various perturbation theories have been developed to evaluate the effective action. In 2023, Wondrak et al. \cite{PhysRevLett.130.221502} used the covariant perturbation theory to compute the scalar particle production rate in a Schwarzschild spacetime, which is analogous to the Schwinger effect. Their result is a little bit different from the standard Hawking radiation, and is thought as a new avenue to the black hole evaporation. Their research has attracted significant attention and comments immediately \cite{Ferreiro:2023jfs,Wondrak:2023hcz,Hertzberg:2023xve,Akhmedov:2024axn}, prompting us to reexamine the relationship between the gravitational Schwinger mechanism and the Hawking effect.

As early as 1999, Parker and Raval \cite{Parker:1999td} derived a minimal effective action with a non-trivial imaginary part using the R-summed Schwinger-DeWitt expansion, and applied it to the study the particle production in cosmology \cite{Parker:2000pr}. The imaginary part of the effective action given in \cite{Parker:1999td} reduces to the main result in \cite{PhysRevLett.130.221502} when the background spacetime is Ricci flat. Meanwhile, Dobado and Maroto \cite{Dobado_1999} showed that, in some cases, the results obtained by the perturbative effective action are consistent with the exact Bogolyubov results. 
Although one gets terrific agreement using the perturbation theory, this consistency seems to be a mere coincidence. In Ref.\cite{Ferreiro:2023jfs}, Ferreiro et al. showed that the results obtained by applying perturbation theory to the Schwinger effect in general do not agree with the exact results. Furthermore, some inconsistencies also arise in cosmology scenario \cite{Ferreiro:2023jfs,Hertzberg:2023xve}. They claim that the perturbation formula in Ref.\cite{Wondrak:2023hcz} is an incomplete expression to account for the particle production \cite{Ferreiro:2023jfs}.

In our work, this tension will be reinforced. We will apply the standard heat kernel method to calculate the effective action in (Anti-) de Sitter spacetime and discuss the spontaneous particle production. Due to the maximal symmetry of (A)dS spacetime, the heat equation is exactly solvable \cite{Fukuma_2013}, allowing us to obtain an exact closed expression for the effective action, which can then be compared with the perturbative results. On the other hand, there is a well-known Gibbons-Hawking radiation with the temperature $T = H/2\pi$ in de Sitter spacetime \cite{Gibbons:1977mu}. Therefore, de Sitter spacetime is an excellent test bed to study the connection between the gravitational Schwinger mechanism and the Hawking effect.

In (Anti-) de Sitter spacetime, the spontaneous particle production is a widely studied yet subtle topic. It is generally believed that particle production does not occur in Anti-de Sitter spacetimes as well as in odd-dimensional de Sitter spacetimes. In even dimensional de Sitter spacetimes, however, the situation becomes controversial. The results of the Bogolyubov transformation between some In- and Out- vacuum states indicates the non-perturbative particle production in dS backgrounds. However, Das et al. \cite{Das_2006} calculated the effective action using the coincidence limit of two-point correlation functions and found no imaginary part. Similarly, based on the regularization of the effective action, Kim \cite{kim2010vacuumstructuresitterspace} observed that the imaginary part of the regularized effective action vanishes. Akhmedov et al. \cite{Akhmedov_2019} pointed out that Ref.\cite{Das_2006} incorrectly employed the In-In correlation functions. But rather, by using the correct In-Out correlation functions, they obtained a non-zero result of the imaginary part of the effective action. 


In our work, the exact imaginary part of the effective action obtained from the heat kernel method is consistent with Akhmedov et al.'s result \cite{Akhmedov_2019} in dS spacetime,  and it vanishes in AdS spacetime as one would expect. However, our work also presents significant differences from the predictions of perturbation theory. Furthermore, we note that, even in even-dimensional dS spacetimes, the sign of the imaginary part of the effective action alternates with dimensions, leading to non-unitary results in certain even dimensions, which is overlooked in Ref.\cite{Akhmedov_2019}. In particular, such non-unitary results may occur in four-dimensional spacetime. This makes the interpretation of particle production in dS spacetime via the gravitational Schwinger mechanism require more judicious considerations. However, we argue that this non-unitary result can be eliminated at the heat kernel level. Moreover, we analyze the reasons for the discrepancy between the exact results and the perturbative predictions in (A)dS spacetime and find that the perturbation theory, in general, fails to predict the correct imaginary part of the effective action.   

This work is arranged as follows: In section \ref{GSM}, we  use the Euclidean heat kernel method to provide an exact closed form of the effective action of a massive scalar field in (A)dS spacetime. In section \ref{Compare}, we compare our results with those obtained from other methods to show the applicability of the Euclidean heat kernel method and the limitations of the perturbation theory. Finally, we draw the conclusions in section \ref{conclusion}.

\section{Gravitational Schwinger mechanism}\label{GSM}
Just as the Schwinger effect in QED describes particle production, a similar mechanism in a gravitational field would be referred to as the gravitational Schwinger mechanism. As previously mentioned, the key to identifying the gravitational Schwinger mechanism is the presence of an imaginary part in the effective action.

In this section, we will concentrate on quantum fields interacting with background or external fields which are not quantized, i.e. scalar fields in curved spacetime. By introducing the gravitational field as a background field, the heat kernel method serves as a powerful tool for computing the effective action. The heat kernel is the fundamental solution to the heat equation associated with the operator governing the field in the curved spacetime. By examining the trace of the heat kernel, which encapsulates the effects of quantum fluctuations in the gravitational field, one can systematically derive the effective action. 

Let us consider a massive scalar field $\phi$, described by the action
\bea{}\label{Action_of_Scalar}
S[\phi,g]=\frac{1}{2}\int \dd^n x \sqrt{|g|} \left(g^{\mu\nu}\partial_{\mu}\phi\partial_{\nu}\phi-m^2\phi^2 \right),
\eea
with the signature of the metric $(+,-,-,-)$. We now perform an analytic continuation to the Euclidean time $\tau=it$. The Euclidean action is then given by
\bea{}\label{EuclideanAction}
S_\EE\left[\phi,g\right]=\frac{1}{i}S\left[\phi,g\right]_{t=-i\tau}=\frac{1}{2}\int \dd ^n x\sqrt{g}\left(g^{\mu\nu}\partial_{\mu}\phi\partial_{\nu}\phi+m^2\phi^2 \right),
\eea
where $g_{\mu\nu}$ is the Euclidean metric with the signature $(+,+,+,+)$. Assuming that the field $\phi$ decays quickly enough at infinity, we can integrate by parts in (\ref{EuclideanAction}) and, omitting the boundary terms, rewrite the action (\ref{EuclideanAction}) in the following convenient form
\bea{}\label{qEuclideanAction}
S_\EE\left[\phi,g\right]=\frac{1}{2}\int \dd ^n x\sqrt{g}\phi\left(-\Box+m^2\right)\phi,
\eea
where 
\bea{}\label{Laplacian}
\Box\phi=\frac{1}{\sqrt{g}}\partial_{\mu}\left[ \sqrt{g}g^{\mu\nu}\partial_{\nu}\phi\right]
\eea
is the covariant Laplace operator. The Euclidean effective action $W_\EE$ is defined as
\bea{}
\exp\left(-W_\EE\left[g\right] \right)=\int \mathcal{D}[\phi] e^{-S_\EE\left[\phi,g\right]}.
\eea
Since the action (\ref{qEuclideanAction}) is quadratic, we can formally rewrite the Euclidean effective action as a functional determinant \cite{Mukhanov:2007zz}
\bea{}
W_\EE\left[g\right] = \frac{1}{2}\ln\det \left(-\Box+m^2\right) =-\frac{1}{2}\int_0^{+\infty}\frac{\dd s}{s}e^{-m^2 s}\tr e^{-s\left(-\Box\right)}.
\eea
In the second equality, we have followed Schwinger's trick, where the parameter $s$ is known as Schwinger proper time \cite{Schwartz_2013,alvarez2022CTinQFT}. The trace here encompasses not only discrete indices, but also includes a spacetime integral. For a scalar field, this trace consists solely of the spacetime integral. Since $m^2$ is a constant, it can be factored out of the trace. Here $m^2$ should be understood to be $m^2 - i\epsilon$ with $\epsilon$ a positive infinitesimal for a particular choice of Feynman prescription.

Now we can define the heat kernel associated to the differential  operator $-\Box$ as the operator
\bea{}
K(s)\equiv e^{-s\left(-\Box\right)},
\eea
where the kernel obeys the heat equation
\bea{}
\left[\frac{\partial}{\partial s}+\left(-\Box\right) \right]K(s)=0.
\eea
We can use Dirac notation to express it more explicitly in the coordinate representation. $K(x,y;s)=\left \langle y |K(s) | x  \right \rangle $ satisfies
\bea{}\label{HeatEq}
\left[\frac{\partial}{\partial s}+\left(-\Box\right) \right]K(x,y;s)=0,
\eea
with the boundary condition
\bea{}
K(x,y;0)=\frac{\delta^n(x-y)}{\sqrt{g}}.
\eea
The trace of the heat kernel, which does not depend on the choice of the basis, can be expressed as 
\bea{}
\tr e^{-s\left(-\Box\right)} =\int \dd ^n x \sqrt{g} K(x,x;s).
\eea

Ultimately, we convert the calculation of the effective action into the problem of solving the heat equation (\ref{HeatEq}). Once we obtain the solution to the heat equation, the effective action can, in principle, be immediately computed using the following expression
\bea{}\label{Weff}
W_\EE\left[g\right] = -\frac{1}{2}\int \dd^n x \sqrt{g}\int_0^{+\infty}\frac{\dd s}{s}e^{-m^2 s}K(x,x;s).
\eea
In general spacetime, solving the heat equation is challenging and often requires perturbative techniques to extract limited information \cite{Vassilevich_2003}. Fortunately, in (A)dS spacetime we will discuss, the heat equation is exactly solvable, allowing us to obtain the complete effective action.

In physics, (A)dS spacetime is a solution to the Einstein field equations with a positive (or negative) cosmological constant. AdS spacetime is widely discussed in holography, while dS spacetime is applied in cosmological research. Both are maximally symmetric spacetimes, and their Riemann curvature tensor can be expressed as
\bea{}\label{Rabcd}
R_{\rho\sigma\mu\nu}=\frac{R}{n(n-1)}\left(g_{\rho\mu}g_{\sigma\nu}-g_{\rho\nu}g_{\sigma\mu} \right).
\eea
We define the (A)dS radius $a$ with $\frac{R}{n(n-1)}=\frac{1}{a^2}>0$ for dS and $\frac{R}{n(n-1)}=-\frac{1}{a^2}<0$ for AdS. After performing a Wick rotation to Euclidean signature, the geometry of dS spacetime corresponds to a sphere, while the geometry of AdS spacetime corresponds to a hyperbolic space. The formulas for sphere can be obtained from the formulas for hyperbolic space by replacing $a\to ia$ \cite{Heatkernel_2015}.

We now investigate the heat kernel of the Laplacian on $n$-dimensional hyperbolic space $H^n$ (EAdS) and sphere $S^n$ (EdS). We will carry out the calculations in $H^n$ and then naively perform analytic continuation back to $S^n$. The key observation is that because of the invariance properties the heat kernel $K(x,y;s)=K(r;s)$ can only depend on the geodesic distance $r$ between the points $x$ and $y$. Therefore, the Laplacian (\ref{Laplacian}) in $H^n$ simplifies and becomes an ordinary differential operator, see Ref.\cite{Heatkernel_2015} for detail,
\bea{}
\Box_{H^n}=\partial_r^2+\frac{(n-1)}{a}\coth( r/a )\partial_r.
\eea
This operator simplifies further by introducing new variables $z=\cosh(r/a), \tau=s/a^2$ 
\bea{}
\Box_{H^n}=\frac{1}{a^2}\left[\left(z^2-1 \right)\partial_z^2+n z\partial_z \right],
\eea
and therefore the heat equation (\ref{HeatEq}) simplifies to
\bea{}\label{HKeqEAdS}
\left[\partial_\tau-\left(z^2-1 \right)\partial_z^2-n z\partial_z\right]K_{H^n}(z; \tau)=0.
\eea
The heat kernel satisfies the following useful recurrence relation
\bea{}\label{RecRofHKEAdS}
K_{H^{n+2}}(z;\tau)=-\frac{1}{2\pi a^2}e^{-n \tau}\partial_{z}K_{H^n}(z; \tau).
\eea
Since we are interested in the heat kernel diagonal, we can take the coincidence limit for the heat kernel equation (\ref{HKeqEAdS}) and the recurrence relation (\ref{RecRofHKEAdS}) respectively, i.e., set $z=1$, and joining these two equations we can obtain the recurrence relation for the heat kernel diagonal \cite{yu2018heatkernelrecurrencespace}
\bea{}
K_{H^{n+2}}(x,x;\tau)=-\frac{1}{2\pi  n a^2} e^{-n \tau} \partial_{\tau}K_{H^n}(x,x;\tau).
\eea
Changing the variables back to Schwinger proper time $s$, we obtain the standard form of the recurrence relation for the heat kernel diagonal
\bea{}\label{RecRofHKD}
K_{H^{n+2}}(x,x;s)=-\frac{1}{2\pi n} e^{-\frac{n}{a^2} s } \partial_{s}K_{H^n}(x,x;s).
\eea
The recurrence relations above indicate that we only need to know the expression for the heat kernel in one- and two-dimensional hyperbolic space, which is sufficient to obtain the heat kernel in hyperbolic space of arbitrary dimension. This will greatly simplify our calculations.

For $n=1$ the Laplacian is simply $\partial_r^2$ both in hyperbolic space and sphere. Therefore, the corresponding heat kernel is simply the one-dimensional Euclidean one
\bea{}
K_{\mathbb{R}}(r;s)=\left( 4\pi s\right)^{-1/2}\exp\left(-\frac{r^2}{4s} \right)
\eea
and its diagonal 
\bea{}\label{1DHKD}
K_{\mathbb{R}}(x,x;s)=\left( 4\pi s\right)^{-1/2}
\eea
is a simple power function. This fact leads to the effective action being purely real in any odd dimension.

As a quick check, let us examine the effective action of heavy scalar particles in three-dimensional (A)dS spacetime. Applying the recurrence relation (\ref{RecRofHKD}) to the one-dimensional heat kernel diagonal (\ref{1DHKD}), we obtain the heat kernel diagonal in three-dimensional spacetime
\bea{}
K_{H^3}(x,x;s)=\left( 4\pi s\right)^{-3/2}\exp\left(-\frac{s}{a^2} \right),
\eea
therefore the Euclidean effective action in three-dimensional AdS spacetime given by (\ref{Weff}) reads
\bea{}\label{WEAdS3}
W_\EE^{\text{AdS}_3}=-\frac{1}{2}\int \dd^3 x \sqrt{g}\int_0^{+\infty}\frac{\dd s}{s}\frac{1}{\left(4\pi s \right)^{3/2}}e^{-\left(m^2+\frac{1}{a^2}\right) s}.
\eea
In this form, the effective action is more obviously real. The UV divergence of the integral at $s\to 0$ can be unsettling, finally, the effective action should be renormalized. 
Here, we follow the zeta function regularization procedure \cite{Mukhanov:2007zz}. First, introduce a small parameter $z$ and define the zeta function
\bea{}
\zeta(z)=\frac{1}{\Gamma(z)}\int_0^{+\infty}\frac{\dd s}{s^{1-z}}e^{-m^2 s}\tr K(s).
\eea
Then the finite part of the effective action can be expressed as the derivative of the zeta function at $z = 0$
\bea{}
W_\EE[g]=-\frac{1}{2}\frac{\dd \zeta(z)}{\dd z} \Big |_{z=0} .
\eea
Perform this procedure for (\ref{WEAdS3}), we obtain the renormalized effective action 
\bea{}
W_\EE^{\text{AdS}_3}=-\frac{1}{12\pi}\int \dd^3 x \sqrt{g} \left(m^2+\frac{1}{a^2}\right)^{3/2}.
\eea
By substituting $a→ia$, we obtain the renormalized effective action in dS$_3$ spacetime
\bea{}\label{WE_dS3}
W_\EE^{\text{dS}_3}=-\frac{1}{12\pi}\int \dd^3 x \sqrt{g} \left(m^2-\frac{1}{a^2}\right)^{3/2},
\eea
where the heavy-field condition $m^2-\frac{(n-1)^2}{4a^2}\ge 0$ guarantees convergence of the integral in the dS spacetime and the effective action is purely real.

This calculation can be continued to any odd (A)dS dimensional spacetime by recursive relations. Since the heat kernel is real and does not have any singularities in the positive $s$-axis in both AdS and dS spacetimes, there is no imaginary part of the effective action. This means that particle (at least heavy particle) production does not occur in odd dimensional (A)dS spacetimes, which is consistent with known results \cite{kim2010vacuumstructuresitterspace,Akhmedov_2019}.

Next, our interest is focused on the even dimensional (A)dS spacetime, in which there may be features of particle production. We will start from the heat kernel in $H^2$ and then use recurrence relations and analytic continuation to obtain the effective action in even-dimensional (in particular, four-dimensional) (A)dS spacetimes.

The heat kernel diagonal in $H^2$ takes the form, see Ref.\cite{Heatkernel_2015} for detail,
\bea{}
K_{H^2}(x,x;s)=\frac{1}{4\pi s} \exp\left(-\frac{s}{4a^2} \right)\int_{-\infty}^{+\infty}\frac{\dd\omega}{\sqrt{4\pi}}\exp\left(-\frac{\omega^2}{4} \right)\frac{\omega \sqrt{s}/\left(2a\right)}{\sinh\left[\omega\sqrt{s}/\left(2a\right) \right]},
\eea
which is more complicated than those in odd dimension. It is observed that for positive $s$, the heat kernel diagonal is real and has no singularities, which implies that the effective action in AdS$_2$ is also real. According to the recurrence relation, more generally, the effective action is real in even-dimensional AdS spacetimes, just as in odd-dimensional spacetimes, indicating the absence of particle production. By replacing $a\to ia$, we get the heat kernel diagonal of $S^2$
\bea{}\label{HeatKernelDS2}
K_{S^2}(x,x;s)=\frac{1}{4\pi s} \exp\left(\frac{s}{4a^2} \right)\int_{-\infty}^{+\infty}\frac{\dd\omega}{\sqrt{4\pi}}\exp\left(-\frac{\omega^2}{4} \right)\frac{\omega \sqrt{s}/\left(2a\right)}{\sin\left[\omega\sqrt{s}/\left(2a\right) \right]}.
\eea
We do find that the integrand has poles when $\omega$ equals to $\omega_n=2a \pi  n/\sqrt{s}, n=\pm 1,\pm 2,\cdots$ for $s>0$, which would result in the heat kernel diagonal would have an imaginary part. A convenient way to extract the imaginary part of the heat kernel diagonal in $S^2$ is to first compute the convergent integral in $H^2$ and then perform analytical continuation on the result. 

Recalling the Mittag-Leffler's theorem, we can expand the integrand into a series
\bea{}
\frac{1}{\sinh(z)}=\frac{1}{z}+2z\sum_{n=1}^{\infty}\frac{(-1)^n}{z^2+n^2\pi^2}.
\eea
The integral then takes the form
\begin{align}
I_{H^2}(\chi)=\int_{-\infty}^{+\infty}\frac{d\omega}{\sqrt{4\pi}}\exp\left( -\frac{\omega^2}{4}\right)\left(1+2\omega^2\chi^2\sum_{n=1}^{\infty} \frac{(-1)^n}{\omega^2\chi^2+n^2\pi^2}\right),
\end{align}
where $\chi = \sqrt{s}/(2a)$ for simplicity. Integrating term by term, we obtain
\bea{}
I_{H^2}(\chi)=1+\sum_{n=1}^{\infty}(-1)^n\left[2 +\frac{\pi ^{3/2} }{\chi} n e^{\frac{\pi ^2 n^2}{4 \chi ^2}} \text{erf}\left(\frac{\pi  n}{2 \chi }\right)-\frac{\pi ^{3/2} }{\chi} n e^{\frac{\pi ^2 n^2}{4 \chi ^2}}\right]
\eea
for $\chi>0$ and the $\text{erf}(z)$ denotes the error function. By performing analytic continuation $a\to ia$, which corresponds to $\chi\to -i\chi$, we can obtain the result for the integral in $S^2$
\bea{}
I_{S^2}(\chi)=1+\sum_{n=1}^{\infty}(-1)^n\left[2 -\frac{\pi ^{3/2} }{\chi} n e^{-\frac{\pi ^2 n^2}{4 \chi ^2}} \text{erfi}\left(\frac{\pi  n}{2 \chi }\right)-i\frac{\pi ^{3/2} }{\chi} n e^{-\frac{\pi ^2 n^2}{4 \chi ^2}}\right],
\eea
where $\text{erfi}(z)=\text{erf}(iz)/i $ is real for real $z$. We find an imaginary part in the result, originating from the standalone $1/\chi$ factor. Thus there is an imaginary part of the heat kernel diagonal in $S^2$
\bea{}\label{ImKS2xxs}
\im K_{S^2}(x,x;s)=-\frac{1}{2 s}\frac{a\sqrt{\pi}}{\sqrt{s}} \exp\left(\frac{s}{4a^2} \right)\left[\sum_{n=1}^{\infty}(-1)^n n e^{-\frac{a^2\pi^2n^2}{s}} \right].
\eea
Substituting the imaginary part of the heat kernel diagonal into the expression for the effective action (\ref{Weff}), we get the imaginary part of the effective action in dS$_2$. Define $\mu^2 = m^2-\frac{(D-1)^2}{4a^2} \ge 0$ and assume $\mu>0$ for simplicity (here $D=2$ is the dimension of the spacetime), we get 
\begin{align}
\im W_{\EE}^{\text{dS}_2} &=-\frac{1}{2}\int \dd^2x \sqrt{g} \int_0^{+\infty}\frac{\dd s}{s}e^{-m^2s}  \im K_{S^2}(x,x;s) \\
&=\int \dd^2x \sqrt{g} \left[\sum_{n=1}^{\infty}\frac{(-1)^n e^{-2 \pi  a \mu  n} (2 \pi  a \mu  n+1)}{8 \pi ^2 a^2 n^2}\right] \\
&=-\frac{1}{8\pi^2 a^2} \int \dd^2x \sqrt{g} \left[2 \pi  a \mu  \log \left(e^{-2 \pi  a \mu }+1\right)-\text{Li}_2\left(-e^{-2 a \pi  \mu }\right) \right],  \label{ImWEdS2}
\end{align}
up to an overall minus sign from the Euclidean signature and $\text{Li}_2(x)$ is the dilogarithm function. Applying the Wick rotation back to the Lorentzian signature, we obtain the imaginary part of Lorentzian effective action in dS$_2$
\bea{}\label{ImWdS2}
\im W_{\LL}^{\text{dS}_2} = \frac{1}{8\pi^2 a^2} \int \dd^2x \sqrt{|g|} \left[2 \pi  a \mu  \log \left(e^{-2 \pi  a \mu }+1\right)-\text{Li}_2\left(-e^{-2 a \pi  \mu }\right) \right] > 0,
\eea
which implies that particle production via the gravitational Schwinger mechanism does occur in dS$_2$ spacetime. 

The expression for the imaginary part of the effective action (\ref{ImWdS2}) is a bit complicated, so let us take some limits to see the physical meaning more explicitly. In the first limit, let us take $\mu \to 0$, which corresponds to the lower bound of the principal series in dS,
\bea{}\label{ImWdS2mu0}
\im W_{\LL}^{\text{dS}_2}(\mu\to 0) = \int \dd^2x \sqrt{|g|} \left( \frac{1}{96a^2} \right).
\eea
This corresponds to the residue of the integrand in the $s$-integral of effective action (\ref{Weff}) at $s=0$, which may be reconstructed from perturbation theory.

The other limit we will discuss is called the large-mass/weak-curvature limit with $m \gg 1/a = H $. In this limit, (\ref{ImWdS2}) simplifies to
\bea{}
\im W_{\LL}^{\text{dS}_2}(m\gg 1/a)=\int \dd^2x \sqrt{|g|} \left( \frac{m}{4\pi a} e^{-2\pi a m} \right).
\eea
The factor $e^{-2\pi a m}$ reminds us of the thermal distribution with a temperature of 
\bea{}
T=\frac{1}{2\pi a}=\frac{H}{2\pi},
\eea
which is nothing but Gibbons-Hawking temperature in dS spacetime. $H$ is the Hubble's constant.

So far, everything is consistent with the well-knwon results in (A)dS spacetime. Let us continue to examine the four-dimensional case. Performing the recurrence relation of dS version (replace $a\to ia$ in (\ref{RecRofHKD})) on (\ref{ImKS2xxs}), we can obtain the imaginary part of the heat kernel diagonal in $S^4$,
\begin{align}
\im K_{S^4}(x,x;s)=\frac{\exp\left(\frac{9 s}{4 a^2}\right)}{32 \sqrt{\pi} a s^{7/2}}\sum_{n=1}^{\infty}(-1)^{n}n\left(4 \pi ^2 a^4 n^2-6 a^2 s+s^2\right) e^{-\frac{\pi ^2 a^2 n^2}{s}} .
\end{align}
Substituting the above expression into the effective action formula (\ref{Weff}), we obtain the imaginary part of the effective action in dS$_4$. In terms of $\mu^2 = m^2 - \frac{9}{4a^2}$, the result reads,
\begin{align}
\im W_{\EE}^{\text{dS}_4}=&\frac{1}{128 \pi ^5 a^4}\int \dd^4x\sqrt{g}\left[2 \pi ^3 a \mu  \left(4 a^2 \mu ^2+1\right) \log \left(e^{-2 \pi  a \mu }+1\right)-\right.\notag\\
&\left.\pi ^2 \left(12 a^2 \mu ^2+1\right) \text{Li}_2\left(-e^{-2 a \pi  \mu }\right)-12 \pi  a \mu  \text{Li}_3\left(-e^{-2 a \pi  \mu }\right)-6 \text{Li}_4\left(-e^{-2 a \pi  \mu }\right)\right].
\end{align}
After a Wick rotation to Lorentzian signature, we can obtain the imaginary part of the effective action in dS$_4$,
\begin{align}\label{ImWLdS4}
\im W_{\LL}^{\text{dS}_4}=&-\frac{1}{128 \pi ^5 a^4}\int \dd^4x\sqrt{|g|}\left[2 \pi ^3 a \mu  \left(4 a^2 \mu ^2+1\right) \log \left(e^{-2 \pi  a \mu }+1\right)-\right.\notag\\
&\left.\pi ^2 \left(12 a^2 \mu ^2+1\right) \text{Li}_2\left(-e^{-2 a \pi  \mu }\right)-12 \pi  a \mu  \text{Li}_3\left(-e^{-2 a \pi  \mu }\right)-6 \text{Li}_4\left(-e^{-2 a \pi  \mu }\right)\right].
\end{align}
It is observed that the imaginary part of the effective action contains a physically dubious minus sign, i.e. $\im W_{\LL}^{\text{dS}_4}<0$ (the green line in Fig.\ref{fig.ImLdS4}). Recall that the imaginary part of the effective action corresponds to the probability of particle production and thus its value should lie within the range of $0$ to $1$. A negative "probability" physically undermines unitarity. This indeed occurs in dS$_4$ spacetime if we perform the analytic continuation in a manner consistent with that used for dS$_2$. Our results are consistent with Ref.\cite{Akhmedov_2019}, where this unusual negative sign is explicitly present but was overlooked.

Although there are some issues with interpreting particle production using the imaginary part of the effective action in dS$_4$ spacetime, we still proceed with a limiting analysis. When $\mu\to 0$, 
\bea{}\label{ImWdS4mu0}
\im W_{\LL}^{\text{dS}_4}(\mu\to 0)=\frac{1}{64\pi}\int \dd^4x \sqrt{|g|}\left(-\frac{17}{240a^4}\right).
\eea
This expression matches the perturbative results in Ref.\cite{Parker:1999td,Parker:2000pr,Parker_Toms_2009} when the step function $\theta(x)$ is dropped in their perturbation formula.

In the large mass limit, the expression reduces approximately to
\bea{}
\im W_{\LL}^{\text{dS}_4}(m\gg 1/a) = \int \dd^4x \sqrt{|g|} \left(-\frac{m^3}{16\pi^2 a}e^{-2\pi a m } \right).
\eea
Again, we find the Gibbons-Hwaking temperature $T=1/(2\pi a)=H/(2\pi)$.

\section{Comparisons with other methods}\label{Compare}
The (A)dS spacetime furnishes pivotal exact solutions instrumental in the exploration of non-perturbative gravitational particle production. In the previous section, we performed an exact calculation of the imaginary part of the one-loop effective action for a massive scalar field in (A)dS spacetime using standard heat kernel methods.

We used the Euclidean approach, transforming spacetime into the Euclidean domain to simplify the computation of the effective action and gain valuable insights into particle production in (A)dS spacetimes. In contrast, the Lorentzian method maintains a real-time description of spacetime, which may lead to subtle differences in interpreting results. Therefore, in this section, we will first validate the consistency between the effective actions obtained via the Euclidean technique and those from established Lorentzian frameworks \cite{Akhmedov_2019}, thereby ensuring that our results faithfully represent physical reality.

Next, we will juxtapose our exact results with those derived from perturbation theory \cite{Parker:1999td,Wondrak:2023hcz}. Perturbative calculations offer approximate solutions by expanding around a solvable background (e.g. nearly flat domain) and considering small deviations. However, these approximations may not capture all the intricate details of the particle production \cite{Ferreiro:2023jfs}. By comparing our exact results with perturbative outcomes, we seek to reveal any significant discrepancies and refine our understanding of the applicability and limitations of perturbative techniques.\\

$\bullet$ \underline{Euclidean \textit{v.s.} Lorentzian} : Consistency

The effective action can also be obtained through the two-point function (Feynman propagator). Recalling the definition of the effective action (\ref{defWeff1}) and substituting the action for a massive scalar field (\ref{Action_of_Scalar}), 
\bea{}
W_{\LL}=-i\log\int \mathcal{D}\phi e^{iS[\phi,g]},
\eea
where
\bea{}
S[\phi,g]=\frac{1}{2}\int \dd^n x \sqrt{|g|} \left(g^{\mu\nu}\partial_{\mu}\phi\partial_{\nu}\phi-m^2\phi^2 \right).
\eea
We can derive the effective action using the trick of differentiating with respect to $m^2$ and then integrating. 
\bea{}
\frac{\partial W_{\LL}}{\partial m^2}=-\frac{1}{2}\int \dd^n x\sqrt{|g|} \frac{\int \mathcal{D}[\phi] \phi(x)\phi(x)e^{i S[\phi]} }{\int \mathcal{D}[\phi] e^{i S[g,\phi]} }=-\frac{1}{2}\int \dd^n x\sqrt{|g|} G_F(x,x),
\eea
where the Feynman propagator $G_F(x,y)$ satisfies
\bea{}
\left(\Box+m^2\right)G_{F}(x,y)=-i\frac{\delta^n(x-y)}{\sqrt{|g(x)|}}.
\eea
This allows to express the effective action via Feynman propagator
\bea{}\label{WformGF}
W_{\LL}=-\frac{1}{2}\int \dd^n x\sqrt{|g|} \int _{+\infty}^{m^2}\dd \bar{m}^2 G_F(x,x).
\eea
Here, by setting the lower limit of the integral over $m^2$ to positive infinity, we provide an interpretation from the perspective of particle production: since moving and creating a particle with infinite mass require infinite energy, we expect that as $m\to \infty$, the imaginary part of the effective action (if it exists) must vanish.

In de Sitter spacetime, the Feynman propagator of the In-Out vacuum is already known as (see Ref.\cite{Akhmedov_2019,Fukuma_2013} for details),
\bea{}\label{dSFeynmanPropagator}
G_F(u)=\frac{(-a)^{-n+2}}{(2\pi)^{\frac{n}{2}}}\left(u^2-1 \right)^{-\frac{n-2}{4}}Q^{\frac{n-2}{2}}_{-\frac{1}{2}+i\nu}(u),
\eea
where $Q$ is the associated Legendre function of the second kind, $u=\cos\left(r/a \right) -i\epsilon $ is the hyperbolic distance with an $i\epsilon$ shift and $\nu=\mu a=\sqrt{m^2a^2-\frac{(n-1)^2}{4}}$ is a dimensionless parameter.

To obtain the effective action, we need to compute the coincident limit of the Feynman propagator. As in Ref.\cite{Akhmedov_2019}, we insert the integral representation of the Legendre function of the second kind 
\bea{}
Q^{\mu}_{\nu}(z)=\frac{e^{\mu\pi i }}{2^{\nu+1}}\frac{\Gamma(\nu+\mu+1)}{\Gamma(\nu+1)}(z^2-1)^{\mu/2}\int_{-1}^{1}(1-t^2)^{\nu}(z-t)^{-\nu-\mu-1}dt
\eea
into (\ref{dSFeynmanPropagator}), and then set $u=1$
\begin{align}
G_F(u=1)=&2^{-i \nu -\frac{n}{2}-\frac{1}{2}}\pi ^{\frac{1}{2}-\frac{n}{2}} e^{\frac{1}{2} i \pi  (n-2)} (-a)^{-n+2} \frac{\Gamma \left(i \nu +\frac{n}{2}-\frac{1}{2}\right)}{\Gamma (i \nu+1 )} \times \notag\\
&\, _2F_1\left(\frac{1}{4} (2 i \nu +n-1),\frac{1}{4} (2 i \nu +n+1);i \nu+1 ;1\right),
\end{align}
where $_2F_1$ is the hypergeometric function. Using the Gamma function representation of the hypergeometric function evaluated at $z=1$
\bea{}
\, _2F_1(\alpha,\beta,\gamma,1)=\frac{\Gamma(\gamma)\Gamma(\gamma-\alpha-\beta)}{\Gamma(\gamma-\alpha)\Gamma(\gamma-\beta)} \ , \ \text{Re}(\gamma-\alpha-\beta)>0,
\eea
the expression simplifies to 
\bea{}\label{GF(z=1)}
G_F(u=1)=\frac{e^{-i\pi\frac{n-2}{2}}}{(4\pi)^{\frac{n}{2}}}a^{2-n}\Gamma(1-\frac{n}{2})\frac{\Gamma\left(i\nu+\frac{n-1}{2} \right)}{\Gamma\left(i\nu-\frac{n-3}{2} \right)},
\eea
which is consistent with the result in Ref.\cite{Akhmedov_2019}. In fact, the above expression holds only for $n\le 1$; for the other dimensions of interest, it should be regarded as the result of analytic continuation.

In odd dimensions, for $\nu\ge 0$, $G_F(u=1)$ is purely real and free of any singularities. Thus, the effective action does not exhibit an imaginary part, which is consistent with previous conclusions. For the real part of the effective action, naively substituting (\ref{GF(z=1)}) into the formula of the effective action (\ref{WformGF}), as usual, leads to divergences. We will show that, specifically for three dimensional dS spacetime as an example, the finite part of the expression agrees with the result obtained using the Euclidean heat kernel method. For $n=3$, (\ref{GF(z=1)}) reduces to
\bea{}
G^{\text{dS}_3}_F(x,x)=-\frac{\nu }{4 \pi  a}=-\frac{1}{4\pi a}\sqrt{m^2a^2-1}.
\eea
Substituting it into the formula (\ref{WformGF}), we have
\begin{align}
W_{\LL}^{\text{dS}_3}&=-\frac{1}{2}\int \dd^3 x\sqrt{|g|} \int _{\Lambda^2}^{m^2}\dd\bar{m}^2 G^{\text{dS}_3}_F(x,x) \\
&=\frac{1}{12\pi}\int \dd^3 x\sqrt{|g|} \left[\left(m^2-\frac{1}{a^2}  \right)^{3/2} -\left(\Lambda^2-\frac{1}{a^2}  \right)^{3/2}\right] ,
\end{align}
where $\Lambda$ is a regularization parameter. Expanding the integrand in a Taylor series as $\Lambda \to +\infty$ and getting rid of the divergent terms, we get
\bea{}
W_{\LL}^{\text{dS}_3}=\frac{1}{12\pi}\int \dd^3 x\sqrt{|g|} \left(m^2-\frac{1}{a^2}  \right)^{3/2} ,
\eea
which is consistent with the previous result (\ref{WE_dS3}) and the minus sign disappears due to the Lorentzian signature.

In even-dimensional spacetime, the Gamma function  $\Gamma(1-n/2)$ being at poles necessitates the renormalization of (\ref{GF(z=1)}). We achieve this through dimensional regularization and minimal subtraction in the $\overline{MS}$ scheme. In general, the renormalized Feynman propagator $G_F^{\text{ren}}(x,x)$ can be expressed in terms of Gamma function $\Gamma(z)$ and digamma function $\psi(z)$,
\begin{align}
G_F^{\text{ren}}(x,x)=\frac{a^{-n+2}}{(4\pi)^{\frac{n}{2}}\Gamma\left( \frac{n}{2} \right)}\frac{\Gamma\left(i\nu+\frac{n-1}{2} \right)}{\Gamma\left(i\nu-\frac{n-3}{2} \right)}\left[ i\pi - \psi\left(i\nu+\frac{n-1}{2} \right)-\psi\left(i\nu-\frac{n-3}{2} \right)\right].
\end{align}
It has a non-trivial imaginary part and can be derived into a compact expression using the reflection formula of the Gamma function $\Gamma(z)\Gamma(1-z)=\frac{\pi}{\sin(\pi z)}$ and properties of the digamma function $\im\psi(\frac{1}{2}+i x)= \frac{\pi}{2}\tanh(\pi x)$ for real $x$:
\begin{align}
\im G_F^{\text{ren}}(x,x) =& \frac{a^{-n+2}}{(4\pi)^{\frac{n}{2}}\Gamma\left( \frac{n}{2} \right)}\Big|\Gamma\left(i\nu+\frac{n-1}{2} \right)\Big|^2 \frac{\sin\left[ \pi\left( i\nu-\frac{n-3}{2} \right) \right]}{\pi}\left[\pi - \pi\tanh(\pi\nu) \right] \\
=&-\frac{a^{-n+2}}{(4\pi)^{\frac{n}{2}}\Gamma\left( \frac{n}{2} \right)}\Big|\Gamma\left(i\nu+\frac{n-1}{2} \right)\Big|^2  (-1)^{\frac{n}{2}}\cosh(\pi\nu)\left[1 - \tanh(\pi\nu) \right]\\
=&-\frac{(-1)^{\frac{n}{2}}}{(4\pi)^{\frac{n}{2}}\Gamma\left( \frac{n}{2} \right)}a^{-n+2}e^{-\pi\nu}\Big|\Gamma\left(i\nu+\frac{n-1}{2} \right)\Big|^2 
\end{align}
\bea{}\label{ImGFren}\boxed{
\im G_F^{\text{ren}}(x,x)=-\frac{(-1)^{\frac{n}{2}}}{(4\pi)^{\frac{n}{2}}\Gamma\left( \frac{n}{2} \right)}a^{-n+2}e^{-\pi\nu}\Big|\Gamma\left(i\nu+\frac{n-1}{2} \right)\Big|^2     }.
\eea
The result here differs from that in Ref.\cite{Akhmedov_2019} by an additional minus sign, which may suggest that Ref.\cite{Akhmedov_2019} possibly overlooked this detail. The factor $(-1)^{\frac{n}{2}}$ in the expression (\ref{ImGFren}) indicates that the sign of the imaginary part alternates with the dimension, which provides a clue to the underlying cause of the changes in the sign of the imaginary part of the effective action.

We are particularly focused on the two-dimensional and four-dimensional cases. We will next show that direct calculations using Lorentzian signature yield results consistent with those obtained through the Euclidean method, thereby confirming the validity of the Euclidean approach. In dS$_2$, the expression (\ref{ImGFren}) reduces to
\bea{}
\im G_F^{\text{dS}_2}(x,x)=\frac{1}{4}\left[1-\tanh\left(\pi\nu\right) \right]=\frac{1}{4}\left[1-\tanh\left(\pi\sqrt{m^2a^2-\frac{1}{4}}\right) \right].
\eea
Substituting it into the formula (\ref{WformGF}), we can obtain the imaginary part of the effective action in dS$_2$
\begin{align}
\im W_{\LL}^{\text{dS}_2}&=-\frac{1}{2}\int \dd^2 x\sqrt{|g|} \int _{+\infty}^{m^2}\dd\bar{m}^2 \im G^{\text{dS}_2}_F(x,x) \\
&=-\frac{1}{8}\int \dd^2 x\sqrt{|g|} \int _{+\infty}^{m^2}\dd\bar{m}^2 \left[1-\tanh\left(\pi\sqrt{\bar{m}^2a^2-\frac{1}{4}}\right) \right] \\
&=-\frac{1}{8}\int \dd^2 x\sqrt{|g|} \int _{+\infty}^{\nu}\dd\bar{\nu}\frac{2\bar{\nu}}{a^2} \left[1-\tanh\left(\pi\bar{\nu}\right) \right]   \\
&=\frac{1}{8\pi^2 a^2} \int \dd^2x \sqrt{|g|} \left[2 \pi   \nu  \log \left(e^{-2 \pi  \nu }+1\right)-\text{Li}_2\left(-e^{-2  \pi  \nu }\right) \right] .
\end{align}
Recalling $\nu=a \mu $, here we get the same result as (\ref{ImWdS2}). Performing the same calculation in dS$_4$, 
\begin{align}
\im G_F^{\text{dS}_4}(x,x)=-\frac{\left(4 \nu ^2+1\right)\left[1-\tanh (\pi  \nu )\right]}{64 \pi  a^2} =-\frac{m^2a^2-2}{16\pi a^2}\left[1-\tanh\left(\pi\sqrt{m^2a^2-\frac{9}{4}}\right) \right]
\end{align}
we can get
\begin{align}
\im W_{\LL}^{\text{dS}_4}=&-\frac{1}{2}\int \dd^4 x\sqrt{|g|} \int _{+\infty}^{m^2}\dd\bar{m}^2 \im G^{\text{dS}_4}_F(x,x)\\
=&-\frac{1}{2}\int \dd^4 x\sqrt{|g|} \int _{+\infty}^{\nu}\dd\bar{\nu}\frac{2\bar{\nu}}{a^2} \left[-\frac{\left(4 \bar{\nu} ^2+1\right) (1-\tanh (\pi  \bar{\nu} ))}{64 \pi  a^2} \right]\\
=&-\frac{1}{128 \pi ^5 a^4}\int \dd^4x\sqrt{|g|}\left[2 \pi ^3  \nu  \left(4  \nu ^2+1\right) \log \left(e^{-2 \pi   \nu }+1\right)-\right.\notag\\
&\left.\pi ^2 \left(12 \nu ^2+1\right) \text{Li}_2\left(-e^{-2  \pi  \nu }\right)-12 \pi   \nu  \text{Li}_3\left(-e^{-2  \pi  \nu }\right)-6 \text{Li}_4\left(-e^{-2  \pi  \nu }\right)\right].
\end{align}
Once again, we get the same results as before (\ref{ImWLdS4}). This result obtained from Lorentzian signature confirms that the imaginary part of the effective action in dS$_4$ spacetime is negative, which makes it challenging to use this negative imaginary part to explain particle production in a physical context.\\

$\bullet$ \underline{Non-perturbative \textit{v.s.} Perturbative} : Discrepancy

We have demonstrated the applicability of the Euclidean heat kernel method for calculating the effective action in dS spacetime. Now, let us return to the heat kernel itself. Solving the heat equation in general backgrounds is a challenging task; therefore, perturbation theory has been developed to evaluate the effective action. The perturbative methods typically involve expanding the heat kernel as a series in terms of Schwinger proper time (Seeley–DeWitt expansion \cite{Birrell:1982ix}) or spacetime curvature (covariant perturbation theory \cite{Barvinsky:1990up,Vassilevich_2003}), providing consistent results up to total derivative terms at finite orders. Parker, Wondrak, et al. have claimed that finite-order perturbative results can yield information about the non-perturbative imaginary part of the effective action (i.e. particle production) \cite{Parker:1999td,Wondrak:2023hcz}. 

In the following, we will illustrate that, although the finite-order perturbation theory may capture certain features of the full heat kernel, this information is insufficient to conclusively determine the presence of an imaginary part in the actual heat kernel.

The Seeley–DeWitt expansion, also known as adiabatic expansion, is to expand the heat kernel as a power series near the small Schwinger proper time $s$. In this perturbative expansion, the coincidence limit of the heat kernel (i.e. the heat kernel diagonal) can be expressed as follows,
\bea{}
K(x,x,s)=\frac{1}{\left(4\pi s \right)^{\frac{n}{2}}}\left[ a_0(x) +a_1(x)s +a_2(x) s^2+\cdots  \right],
\eea
where the coefficients $a_k(x)$, known as the Seeley–DeWitt coefficients, can be constructed from background geometric quantities. For a scalar fields, these coefficients take the form:
\begin{align}
a_0(x) & = 1,\\
a_1(x) & =\frac{1}{6}R,\\
a_2(x) & =\frac{1}{180}\left( R_{\alpha\beta\gamma\delta}R^{\alpha\beta\gamma\delta} -R_{\alpha\beta}R^{\alpha\beta} \right) + \frac{1}{72}R^2 +\frac{1}{30}\Box R,\\
&\cdots \notag
\end{align}
The $k$-th order term in the curvature is fully contained within the coefficient $a_k$, up to total derivative terms. In this sense, the Seeley–DeWitt expansion can also be viewed as an expansion in terms of the curvature \cite{El-Menoufi:2015cqw}. Furthermore, all terms involving the Ricci scalar $R$ (excluding its derivatives) in the expansion coefficients can be re-summed into a non-perturbative factor $\exp\left(\frac{1}{6}R s \right)$ \cite{Parker:1984dj,Jack:1985mw,Ferreiro:2020uno}. Then the R-summed Schwinger-DeWitt expansion of the heat kernel diagonal can be writen as
\bea{}
K(x,x,s)=\frac{1}{\left(4\pi s \right)^{\frac{n}{2}}}\exp\left(\frac{1}{6}R s \right)\left[ \bar{a}_0(x) +\bar{a}_1(x)s +\bar{a}_2(x) s^2+\cdots  \right]
\eea
with the coefficients $\bar{a}_k(x)$
\begin{align}
\bar{a}_0(x) & = 1,\\
\bar{a}_1(x) & =0,\\
\bar{a}_2(x) & =\frac{1}{180}\left( R_{\alpha\beta\gamma\delta}R^{\alpha\beta\gamma\delta} -R_{\alpha\beta}R^{\alpha\beta} \right)  +\frac{1}{30}\Box R,\\
&\cdots \notag
\end{align}
and the Ricci scalar $R$ (excluding its derivatives) will not appear in the expansion coefficients.

We can quickly verify the validity of these expansions using the heat kernel diagonal (\ref{HeatKernelDS2}) in dS$_2$,
\begin{align}
K_{S^2}(x,x;s)=&\frac{1}{4\pi s}\exp\left(\frac{s}{4a^2} \right)\left(1+\frac{1}{12a^2}s+\frac{7}{480a^4}s^2+\cdots \right)\label{S2PE} \\
=&\frac{1}{4\pi s}\left(1+\frac{1}{3a^2}s+\frac{1}{15a^4}s^2+\cdots \right)\label{S2SDW} \\
=&\frac{1}{4\pi s}\exp\left(\frac{s}{3a^2} \right)\left(1+\frac{1}{90a^4}s^2+\cdots \right).\label{S2RSSDW}
\end{align}
Recall that de Sitter spacetime is a maximally symmetric spacetime, where the Riemann curvature tensor has a simple expression (\ref{Rabcd}). Therefore,
\begin{align}
R_{\alpha\beta\gamma\delta}R^{\alpha\beta\gamma\delta}&=\frac{2R^2}{n(n-1)} ,\\
R_{\alpha\beta}R^{\alpha\beta}&=\frac{R^2}{n},
\end{align}
and $R=\frac{n(n-1)}{a^2}$ is a constant. In the special case of $n=2$, it is straightforward to see that the perturbative results match the exact results e.g.
\begin{align}
\bar{a}_2=&\frac{1}{180}\left( R_{\alpha\beta\gamma\delta}R^{\alpha\beta\gamma\delta} -R_{\alpha\beta}R^{\alpha\beta} \right) = \frac{1}{180}\times\frac{R^2}{2}= \frac{1}{180}\times\frac{2}{a^4}=\frac{1}{90a^4},\\
a_2=&\bar{a}_2+\frac{1}{72}R^2=\frac{1}{90a^4}+\frac{1}{72}\times \frac{4}{a^4}=\frac{1}{15a^4}.
\end{align}
However, it is important to note that the non-perturbative factor in (\ref{S2RSSDW}) obtained through resummation differs from the one present in the exact results (\ref{HeatKernelDS2}) and (\ref{S2PE}), which is one reason why perturbative results may not accurately reflect the true physical behavior.

Using perturbation expansion of the heat kernel diagonal, the effective action (\ref{Weff}) can be formally expressed as:
\bea{}
W_\EE\left[g\right] = -\frac{1}{2}\int \dd^n x \sqrt{g}\int_0^{+\infty}\frac{\dd s}{s}e^{-\left(M^2-i\epsilon\right) s}\frac{1}{(4\pi s)^{\frac{n}{2}}}\left(\sum_{k=0}^{\infty} \tilde{a}_k s^k  \right),
\eea
where $M^2$ denotes all the non-perturbative factors in the exponential (e.g. $M^2=m^2$ in Seeley-DeWitt expansion, $M^2=m^2-\frac{1}{6}R$ in R-summed  Seeley-DeWitt expansion) and $\tilde{a}_k$s represent the corresponding expansion coefficients. Here, we explicitly write down the infinitesimal shift $i\epsilon$. The integral over $s$ diverges for $k\le \frac{n}{2}$, so we perform dimensional regularization
\bea{}
W_\EE\left[g\right] = -\frac{1}{2}\lim_{z\to 0}\int \dd^n x \sqrt{g}\tilde{\mu}^{2z}\int_0^{+\infty}\frac{\dd s}{s^{1-z}}e^{-\left(M^2-i\epsilon\right) s}\frac{1}{(4\pi s)^{\frac{n}{2}}}\left(\sum_{k=0}^{\infty} \tilde{a}_k s^k  \right)
\eea
with a complex parameter $z$ and an arbitrary renormalization mass scale $\tilde{\mu}$ to keep proper physical dimensions. Then we can work out the integral term by term,
\bea{}
W_\EE\left[g\right] = -\frac{1}{2}\frac{1}{(4\pi)^{\frac{n}{2}}}\lim_{z\to 0}\sum_{k=0}^{\infty}\int \dd^n x \sqrt{g}\tilde{\mu}^{2z}\frac{\Gamma\left(k-\frac{n}{2}+z \right)}{\left(M^2-i\epsilon \right)^{k-\frac{n}{2}+z}}\tilde{a}_k.
\eea

The effective action will have an imaginary part if $k-n/2$ takes on negative integer values \cite{Wondrak:2023hcz}. To fully understand this, let us consider the cases $j=k-n/2=-2, -1$, and $0$, which are sufficient to reveal all the imaginary parts predicted by perturbation theory in 2 and 4 dimensions. Define
\bea{}\label{Ij(z)}
I_j(z)=\tilde{\mu}^{2z}\left(M^2-i\epsilon \right)^{-(j+z)}\Gamma\left(j+z \right)
\eea
and we have
\begin{align}
&I_0(z\to 0)=\frac{1}{z}-\gamma_E-\log\left(\frac{M^2}{\tilde{\mu}^2}-i\epsilon \right)+\mathcal{O}(z),\\
&I_{-1}(z\to 0)=-\frac{M^2}{z}-(1-\gamma_E)M^2+M^2\log\left(\frac{M^2}{\tilde{\mu}^2}-i\epsilon \right)+\mathcal{O}(z),\\
&I_{-2}(z\to 0)=\frac{M^4}{2z}+\frac{3-2\gamma_E}{4}M^4-\frac{(M^2)^2}{2}\log\left(\frac{M^2}{\tilde{\mu}^2}-i\epsilon \right)+\mathcal{O}(z).
\end{align}
The imaginary part originates from the branch cut of the logarithm (chosen along the negative real axis as in Ref.\cite{Wondrak:2023hcz}). Since
\begin{align}
\log\left(\frac{M^2}{\tilde{\mu}^2}-i\epsilon \right)&=\log\bigg| \frac{M^2}{\tilde{\mu}^2}-i\epsilon  \bigg|+i\text{Arg}\left(\frac{M^2}{\tilde{\mu}^2}-i\epsilon \right) \\
&=\log\bigg|\frac{M^2}{\tilde{\mu}^2}-i\epsilon  \bigg| -i\pi\theta(-M^2),
\end{align}
where $\theta(x)$ is the Heaviside step function
\begin{align}
\theta(x)=\left\{\begin{matrix}
1 \ , \ x>0 \\
\frac{1}{2}\ , \ x=0 \\
0 \ , \ x<0 
\end{matrix}\right.  
\end{align}
Thus, when taking the limit $z\to0$, the expression (\ref{Ij(z)}) could contain a finite imaginary part that is constrained by the step function
\begin{align}
&\im I_0 = \pi \theta(-M^2),\\
&\im I_{-1}=-\pi M^2 \theta(-M^2),\\
&\im I_{-2}=\frac{\pi}{2} \left(M^2\right)^2 \theta(-M^2) .
\end{align}
Based on this, Ref.{\cite{Parker:1999td,Parker:2000pr,Wondrak:2023hcz}} argue that perturbation theory can provide the correct imaginary part of the effective action and thereby has the capacity to predict particle production. However, we must emphasize that the imaginary part of the effective action obtained from perturbation theory is unreliable. 

Let us continue with the example of (A)dS$_2$ spacetime to clarify our point. For $n=2$, the imaginary part predicted by perturbation theory comes from $j=-1$ and $j=0$
\begin{align}\label{WEn=2p}
\im W_{\EE}^{n=2}=-\frac{1}{8}\int \dd^2x\sqrt{g}\theta(-M^2)\left(- M^2\tilde{a}_0+ \tilde{a}_1 \right).
\end{align}
First, we should point out that the step function $\theta(-M^2)$ in the expression (\ref{WEn=2p}) is misleading. From the perturbative perspective, the step function controlling the presence of the imaginary part is entirely determined by the factors in the exponential. However, different expansion methods (e.g. (\ref{S2PE})-(\ref{S2RSSDW})) result in different exponential factors. Mathematically, we are permitted to incorporate any factor into the exponential, at the cost of correspondingly altering the expansion coefficients $a_k$. As a result, the constraints are somewhat ambiguous. Comparing the perturbative result (\ref{WEn=2p}) with the exact result (\ref{ImWEdS2}) in dS$_2$, the R-summed perturbation theory predicts that the effective action will have an imaginary part only when $M^2=m^2-\frac{1}{3a^2}\le 0$, while the exact result (\ref{ImWEdS2}) shows that the effective action has an imaginary part for all $\mu^2=m^2-\frac{1}{4a^2} \ge 0$. Furthermore, the imaginary part of the effective action is present in the region $\mu^2=m^2-\frac{(n-1)^2}{4a^2}\ge 0$. when $n\ge 4$, however, this feature will not be captured by any perturbative expansions in (\ref{S2PE})-(\ref{S2RSSDW}).

\begin{figure}[h!] 
    \centering 
    \includegraphics[width=0.8\textwidth]{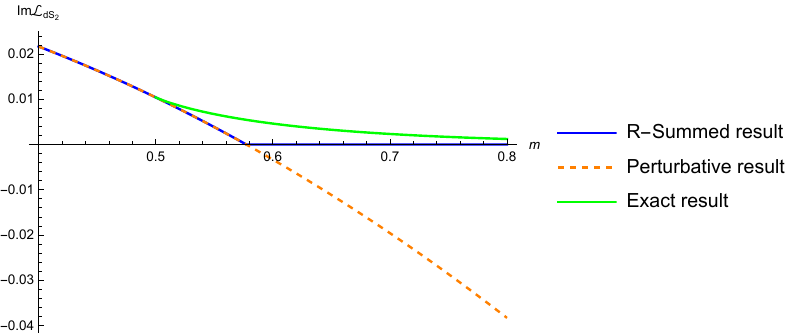} 
    \caption{Comparison of perturbation theory and exact theory predictions for the imaginary part of the Lorentzian effective action in dS$_2$ spacetime with $a=1$: In the region $\frac{1}{4a^2}<m^2<\frac{1}{3a^2}$ where R-summed perturbation theory and exact theory both predict the presence of an imaginary part in the effective action, there are noticeable discrepancies between the perturbative and exact results, except near the special point $\mu^2=m^2-\frac{1}{4a^2}=0$  } 
    \label{fig.ImLdS2} 
\end{figure}

Aside from the step function, let's examine the other components of (\ref{WEn=2p})
\bea{}
\im \mathcal{L}_{\EE}^{n=2}=-\frac{1}{8}\left(- M^2\tilde{a}_0+ \tilde{a}_1 \right)=\frac{1}{8}\left(m^2-\frac{1}{6}R \right).
\eea
Although their expressions might appear to depend on the perturbative expansion method, their values do not.  In fact, they originate from the pole at $s=0$ in the Schwinger representation of the effective action (\ref{Weff}), and their values are proportional to the residue at this pole. In dS$_2$,
\begin{align}\label{ImLpEdS2}
\im\mathcal{L}_{\text{EdS}_2}^{\text{pert}}=\frac{1}{8}\left( m^2 - \frac{1}{3a^2} \right).
\end{align}
The expression (\ref{ImLpEdS2}) shows a significant difference from the exact result (\ref{ImWEdS2}), except for a special case where $\mu^2=m^2-\frac{1}{4a^2}=0$ (See Fig.\ref{fig.ImLdS2} for an intuitive understanding). In this case 
\bea{}
\im\mathcal{L}_{\text{EdS}_2}^{\text{pert}}\left(\mu=0\right)=-\frac{1}{96a^2}.
\eea
which is consistent with the exact result (\ref{ImWdS2mu0}). This result depends on a specific fact: when $\mu^2=0$, the sum of the residues of the integrand with respect to Schwinger proper time $s$ in (\ref{Weff}), taken over all poles in the complex plane (excluding the point at infinity), is zero. This is also why Ref.\cite{Dobado_1999,Wondrak:2023hcz} were able to use perturbation theory to recover the Schwinger effect for massless charged particles in a uniform electric field. In that case, the non-analytical exponential terms in (\ref{SchwingerEffect}) disappear when taking the massless limit $m\to 0$. In fact, in the standard Schwinger effect, only the contributions from the poles on the positive real $s$-axis are considered, while the singularity at $s=0$ has been renormalized away. Therefore, when there is a special relationship between the sum of the residues of the poles on the positive real axis and the residues of the poles that were originally at $s=0$, it is possilble to recover the exact result from perturbation theory. The same is true in the gravitational Schwinger mechanism.

\begin{figure}[h!] 
    \centering 
    \includegraphics[width=0.8\textwidth]{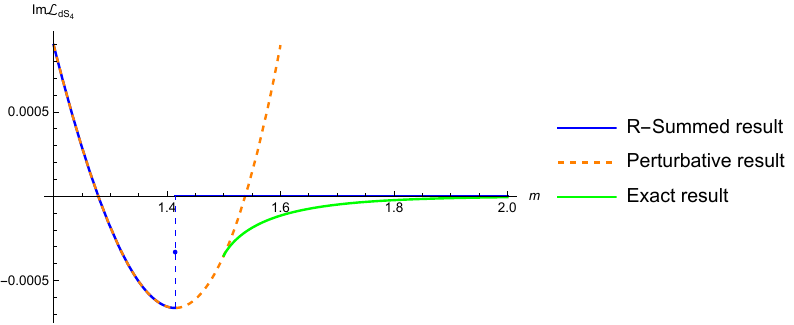} 
    \caption{Comparison of perturbation theory and exact theory predictions for the imaginary part of the Lorentzian effective action in dS$_4$ spacetime with $a=1$: Although there remains a significant discrepancy between the predictions of perturbation theory and the exact result, they still agree at the special point $\mu^2=m^2-\frac{9}{4a^2}=0$, which lies outside the range predicted by the R-summed perturbation theory. } 
    \label{fig.ImLdS4} 
\end{figure}

Let us now quickly examine the case of $n=4$, where perturbation theory gives
\begin{align}
\im \mathcal{L}_\EE^{n=4}&=-\frac{1}{32\pi}\left[\frac{1}{2}\left( M^2\right)^2\tilde{a}_0-M^2\tilde{a}_1+\tilde{a}_2 \right].
\end{align}
When applied to dS$_4$ spacetime, we obtain
\bea{}
\im\mathcal{L}_{\text{EdS}_4}^{\text{pert}}=-\frac{1}{64\pi}\left(m^4-\frac{4m^2}{a^2}+\frac{58}{15a^4} \right).
\eea
Setting $\mu^2=m^2-\frac{9}{4a^2}$, 
\bea{}
\im\mathcal{L}_{\text{EdS}_4}^{\text{pert}}(\mu\to 0)=\frac{1}{64\pi}\times \frac{17}{240a^4}.
\eea
The perturbative result once again coincides with the exact result (\ref{ImWdS4mu0}) up to a minus sign since the Euclidean signature. Fig.\ref{fig.ImLdS4} shows the difference between the imaginary part of the effective action predicted by perturbation theory and the exact results in dS$_4$ spacetime. The only point of agreement lies outside the predictions of the R-summed perturbation theory, further implying that the step function in perturbation theories \cite{Parker:1999td,Parker_Toms_2009,Wondrak:2023hcz} does not accurately characterize the presence of the imaginary part.

We see that the exact results are inconsistent with the perturbative results, except at some special regions for $D=2$ (see Fig.\ref{fig.ImLdS2}). However, these special regions are out of the ranges where the non-zero imaginary part occurs in the perturbative theory when $D\geq4$ (see Fig.\ref{fig.ImLdS4}). This implies that we are nearly unable to capture the non-perturbative information from the perturbative methods.

More unfortunately, the perturbation theory sometimes fails to provide any accurate imaginary part of the effective action. For example, in AdS spacetime, the perturbation theory may predict
\begin{align}
&\im\mathcal{L}_{\text{EAdS}_2}^{\text{pert}}=\frac{1}{8}\left( m^2 + \frac{1}{3a^2} \right),\\
&\im\mathcal{L}_{\text{EAdS}_4}^{\text{pert}}=-\frac{1}{64\pi}\left(m^4+\frac{4m^2}{a^2}+\frac{58}{15a^4} \right).
\end{align}
It is known that at least for $\mu^2= m^2+\frac{(n-1)^2}{4a^2}> 0$, which is above the BF bound, the effective action has no imaginary part, which contrasts with the results from perturbation theory. Specifically, when $\mu^2=0$, the perturbative result involves the residues of poles on the negative  $s$-axis, but these poles are not relevant for the effective action and therefore do not contribute to its imaginary part.

The evidence above continually reinforces the fact that using perturbation theory to assess the imaginary part of the effective action is unconvincing. We have to be judicious about the results from perturbation theory \cite{Parker:1999td,Parker:2000pr,ganguly2024,falcke2024}, and therefore, the existence of the new avenue for black hole evaporation proposed by Wondrak et al.\cite{Wondrak:2023hcz}, based on the perturbative analysis, requires more robust proof.

\section{Conclusions and Discussions}\label{conclusion}
The (A)dS spacetime serves as a crucial framework providing exact solutions that are indispensable for the investigation of non-perturbative gravitational particle production.  In this work,  we have used the exact solutions for the heat equation in E(A)dS space to evaluate the effective action for a massive scalar field. Our approach leverages the recursive relationships of the heat kernel in E(A)dS space, which allows for the computation of heat kernels in arbitrary dimensions.

Our results reveal that in AdS spacetime, the effective action does not exhibit an imaginary part, consistent with previous findings that no spontaneous particle production occurs in AdS spacetime. In odd-dimensional dS spacetime, the effective action also lacks an imaginary component, due to the straightforward generation of the heat kernel from one-dimensional Euclidean space. However, in even-dimensional dS spacetime, the effective action does display a non-trivial imaginary part. We observed that the sign of this imaginary part alternates with the dimensions, with the two-dimensional effective action showing a positive imaginary part while the four-dimensional action exhibiting a negative one. This phenomenon of sign alternation results in the non-unitary outcomes, which complicates the use of the Schwinger mechanism to explain the particle production in certain even-dimensional spacetimes.

We have demonstrated that the imaginary part of the effective action obtained from the Euclidean heat kernel matches the results obtained by Akhmedov et al. \cite{Akhmedov_2019} using the Green's function in the coincidence limit. This confirms the validity of applying the Euclidean method to dS spacetime, despite the non-trivial time evolution present in dS spacetime. On the other hand, a comparison of the exact results from the heat kernel with perturbation theory reveals that, at least in (A)dS spacetime, perturbation theory can hardly capture the correct non-perturbative information. This difficulty arises because Laurent series expansions cannot capture the information from the poles except those at the expansion center (such as the unrenormalized pole in the Schwinger representation at $s=0$), as they will break down when approaching the first singularity. Unless there is a special relationship between other poles and the center pole, such as the sum of the residues of these poles being zero, perturbation theory cannot capture any non-analytic information contributed by other poles. 

We need to stress that the non-unitary results due to the sign alternation in the effective action can, to some extent, be avoided at the heat kernel level. This depends on the way used for analytic continuation from the EAdS to EdS, or the choice of the integration contours in the complex $s$-plane for the Schwinger representation of the effective action (\ref{Weff}). Specifically, in our work, we consistently employed the $a\to ia$ continuation method to extend the heat kernel from the EAdS to EdS, yielding results consistent with previous literature \cite{Akhmedov_2019}. If we appropriately alter the analytic continuation method to $a\to -ia$ for certain even dimensions (e.g. $n=4,8,12,\cdots$), we can eliminate the unphysical signs in the imaginary part of the effective action in those dimensions. If we assume that non-unitary results do not occur in dS spacetimes, then similar cautious methods will be required when evaluating the effective action via the coincidence limit of Green's function in Lorentzian signature.

In summary, although a more judicious consideration is needed to explain the particle production using gravitational Schwinger mechanism in dS spacetime, the significant discrepancies on the imaginary part of the effective action between the exact results (up to a sign) and the perturbative predictions clearly demonstrate that the perturbation theory fails to accurately describe such non-perturbative phenomena. In light of this, previously established results based on the perturbation theory must be re-evaluated to properly account for the non-perturbative effects.

\section*{Acknowledgements}
We are grateful for Prof. E. T. Akhmedov to read the manuscript and give useful suggestions. This work was partially supported by the National Natural Science Foundation of China
(Grants No.12175008).

\bibliographystyle{ieeetr}
\bibliography{ref}

\end{document}